Size distribution and its scaling behavior of InAlAs/AlGaAs quantum dots grown on GaAs by molecular beam epitaxy

X. M. Lu,[1*] M. Koyama,[1] Y. Izumi,[1] Y. Nakata,[2] S. Adachi,[1] and S. Muto[1]

[1]Department of applied physics, Graduate School of Engineering, Hokkaido University, Sapporo, Hokkaido 060-8628, Japan

[2]Fujitsu Laboratories Ltd., 10-1 Morinosato-wakamiya, Atsugi 243-0197, Japan

*xm-lu@eng.hokudai.ac.jp

## ABSTRACT

We studied the size distribution and its scaling behavior of self-assembled InAlAs/AlGaAs quantum dots (QDs) grown on GaAs with the Stranski-Krastanov (SK) mode by molecular beam epitaxy (MBE), at both 480°C and 510°C, as a function of InAlAs coverage. A scaling function of the volume was found for the first time in ternary alloy QDs. The function was similar to that of InAs/GaAs QDs, which agreed with the scaling function for the two-dimensional submonolayer homoepitaxy simulation with a critical island size of $i = 1$. However, a character of $i = 0$ was also found as a tail in the large volume.

## I. INTRODUCTION

Self-assembled semiconductor quantum dots (QDs) have attracted considerable attention owing to their potential significance for quantum information processing such as quantum computing and quantum media conversion.[1] The III-V QDs can be grown easily with the Stranski-Krastanov(SK) mode using molecular beam epitaxy (MBE). Abundant research has been conducted to explain nucleation,[2] atomic interdiffusion,[3-5] size fluctuations, strain anisotropy, and composition of QDs and their effects on photoluminescence.[6-10] Although it is well accepted that QDs are the product of lattice mismatch heteroepitaxial growth process, the precise mechanism of nucleation and size fluctuation of QDs are still the subject for better understanding. To verify whether this common size fluctuation is essential or accidental, we have studied the scaling properties of self-assembled InAs/GaAs QDs in the coverage region befor the density saturation and found that the scaling functions agreed with the homoepitaxy simulation,[11] with a critical island size of $i = 1$ at growth temperatures less than 550°C.[12,13]

The scaling property of a two-dimensional (2D) island size distribution is known for submonolayer coverage. According to the scaling assumption,[14] the island distribution with island size $s$ is given by

$$N_s = \frac{\theta}{\langle s \rangle^2} f\left(s/\langle s \rangle\right). \quad (1)$$

Here, $N_s$ is the density of the islands that contain $s$ atoms, $\theta$ is the fractional surface coverage, $\langle s \rangle$ is the average number of atoms in an island, and $f\left(s/\langle s \rangle\right)$ is the island volume distribution scaling function. The volume distribution scaling Eq. (1) is known to hold for two-dimensional islands of homo- and hetero-submonolayer epitaxy, such as Fe on Fe,[15] InAs on GaAs,[16] as well as for their growth simulations.

Recently it was reported that the growth simulation of the three-dimensional islands gave the same scaling function as the 2D islands.[17]

InAlAs QDs have the advantage of having a luminescence of 750 nm, at which the sensitivity of a Si CCD is maximum. However, there has been no research as far as we know, on the scaling properties of ternary alloy QDs such as the InAlAs QD, which is expected to be more complicated in terms of surface migration and atomic interdiffusion than binary InAs QDs.

In this study, we investigated the size distribution and the scaling behavior of InAlAs/AlGaAs QDs grown at different substrate temperatures and coverages on GaAs with the SK mode by MBE. The size of the QDs was analyzed by atomic force microscopy (AFM) in air.

## II. EXPERIMENTS

Figure 1 shows the structure grown by MBE on semi-insulating GaAs (001) substrates and the result of AFM measurements. The growth conditions and its effect on the size and density of QDs have been described in detail elsewhere.[18] In brief, the $In_{0.7}Al_{0.3}As$ QDs were grown on the $Al_{0.35}Ga_{0.65}As$ layer, with an $As_4$ pressure of $6 \times 10^{-6}$ Torr. Substrate temperature of Ts was calibrated by the oxide desorption of GaAs. The "relative" InAlAs coverage of $t/t_{3D}$ was varied from 1.0 to 1.5 with a growth rate of 0.025 ML/s. Here, $t$ is the growth time, and $t_{3D}$ is the time at 2D-3D transition, which is defined by the transition of the reflection high-energy electron diffraction (RHEED) pattern from streaky to spotty. After the growth of the top InAlAs QDs, the substrate was held for 1 min at Ts before cooling down. The height and diameter of the QDs were analyzed from AFM image using public domain software (ImageJ).[19]

### III. RESULTS AND DISCUSSION

The diameter and height histograms from the QD samples grown at Ts of 480°C and 510°C are shown in Fig. 2. Inset numbers are relative coverage $t/t_{3D}$ of 1.0, 1.2, and 1.4. The average diameter of the InAlAs QDs is nearly the same as that of InAs QDs, but the average height is substantially smaller. It should be noted that the height and diameter determined from the AFM image are not accurate. For example, the tip artifact in the AFM observations was reported to overestimate the diameter by 8 nm in the case of the InAs QDs.[20] However, we believe that this does not change the general tendency of the geometry, and particularly the volume distribution scaling. In fact, excellent agreement with 2D simulation reported in Ref. 12 was obtained by the same method. Besides, it should be noted that the lower aspect ratio (height/diameter) in the InAlAs QDs results in less artifacts of the QD diameter than in the InAs QDs.

As the substrate temperature increased from 480°C to 510°C, both the dot height and the diameter increased (see Fig. 2), but the dot density decreased (see Ref. 18). This tendency is the same as that observed for the InAs QDs.[21-23] In the case of the InAs QDs, this is interpreted as follows: as the temperature increases, the surface migration of the

In atoms is enhanced and the distance between the In atoms that have gathered to form a single dot is increased, resulting in the formation of larger dots with lower densities. The histograms of height and diameter were fitted by Gaussian functions plotted by the continuous curve.

Figure 3 shows the volume histogram of the QDs grown at Ts = 480°C and 510°C. Inset numbers are relative coverage $t/t_{3D}$ of 1.0, 1.2, and 1.4. We assumed that the QD volume is represented by the product $\pi \times (diameter/2)^2 \times height$. In fact, the volume of cap-like dot and cone-shaped dot have prefactors $1/2$ and $1/3$, respectively. However, we note that the scaling function discussed below is independent of the prefactor. At Ts = 480°C, the volume increased sharply when $t/t_{3D}$ changed from 1.0 to 1.2 (see Figs. 3(a) and 3(b)), whereas the volume increased only slightly when $t/t_{3D}$ changed from 1.2 to 1.4 (see Figs. 3(b) and 3(c)). At 510°C, the volume histograms are broad, and especially at $t/t_{3D}$ =1.2 and 1.4 (see Figs. 3(e) and 3(f)), the larger QDs (volume > 16 × $10^3$ nm$^3$) which do not exist in other conditions were observed. The discreteness of the histograms Fig. 3(e) and 3(f) could be simply due to the low density of the QDs.

Figures 4(a) and 4(b) show as references for the average diameter and height as functions of $t/t_{3D}$ from Ref. 18. Figure 4(c) shows the fluctuation of height ($\Delta h/h$) and diameter ($\Delta d/d$), defined by the ratio of the full width at half maximum (FWHM) to the peak value of the Gaussian fit. At 480°C, the fluctuation of height and diameter were almost constant at 30% and 65%, respectively. At 510°C, the fluctuations were scattered over a wide range. However, for both 480°C and 510°C, the volume fluctuation, given by $(\Delta h/h) + 2(\Delta d/d)$, was approximately constant, suggesting the presence of volume scaling function.

Figure 5 shows the scaling plot of dot volume distribution given by Eq. (1). Here, coverage $\theta$ in Eq. (1) was replaced by the effective coverage $\theta_{eff} = \sum sN_s$, obtained by the total QD volume. For comparison, we show the empirical scaling function proposed by J. G. Amar et al.[11] to describe a homoepitaxial growth simulation:

$$f_i(s/\langle s \rangle) = C_i (s/\langle s \rangle)^i e^{-ia_i(s/\langle s \rangle)^{1/a_i}}, \quad (2)$$

where $i$ is the critical island size, or the largest unstable size, defined by that $i + 1$ is the number of atoms needed to form a stable island. $C_i$ and $a_i$ are constants that depend on $i$. The experimental results for 480°C and 510°C are shown in Fig. 5 by solid circles and triangles, respectively. The solid, dashed, and dotted curves are the scaling function $f_i(s/\langle s \rangle)$ of Eq. 2 with $i$ = 1, 2, and 3, respectively. It should be noted that there is no fitting parameter in these curves. Scaling is evident in Fig. 5. As far as we know, this is the first time that a volume scaling function was found for ternary alloy QDs where two kinds of atoms (In and Al) contribute to QDs.

For As rich growth condition, InAs QDs can be classified as one atom model, in which QD formation is determined by one kind of atom In; InAlAs can be classified as two atoms model, in which QD formation is determined by two kinds of atoms: In and Al with quite different surface mobility. Similar to the case of InAs QDs, the InAlAs QDs volume distributions have the better agreement to the function with $i$ = 1 than $i$ = 2 and 3. Agreement of the scaling function to $i$ = 1 suggests that the In adatoms diffuse but dimmers or larger clusters cannot move. In other words, the nucleation sites are formed by the nearest-neighbor collision of group III adatoms. We note here, however, that the physical meaning of the critical island size $i$ obtained by the volume distribution is not straightforward. J. G. Amar *et. al.* predicted that the most accurate method to determine the critical island size is a direct determination of $\chi_1^{'}$ ( $\chi_1^{'} = i/(i+2.5)$ ) describing the flux dependence of the peak island density for 3D islands.[17] Moreover, M. Fanfoni *et. al.*[24] have shown that the volume distribution of the InAs QDs agreed well with the area distribution of the capture zones, or Voronoi cells,[25] which tessellate the surface according to the nearest island. They claimed that the scale invariance seen in the capture zones is more essential than the volumue disutribution. There in fact is an argument to deduce the critical island size from the capture zone scaling fucntion.[26] However, from application point of view, the volume distribution is more important to know, and the argument concerning Eq. (2) can help us to consult simulation results. So far, there is no theory for QDs scaling function even for InAs QDs. We just know empirically that it is similar to 2D simulation as given by J. G. Amar.[11] We use it as a guide here.

Figure 6 shows the scaling plot in Fig. 5 averaged over different $t/t_{3D}$. The solid circles and triangles are for 480°C and 510°C, respectively. Compared to the scaling of InAs/GaAs QDs,[12] the agreement to critical island size $i$ = 1 is not as good. Especially, the current scaling function seems to have a longtail beyond $s/\langle s \rangle = 2$. This is characteristic of $i$ = 0 simulation in which spontaneous nucleation or freezing of monomers occurred due to impurities or surfactants.[11] Since there is no simple analytic form for $f_0(s/\langle s \rangle)$ of $i = 0$, we simply compared our data with the dashed curve which is $i = 0$ simulation results extracted from Ref. 11. The comparison of our data with the $f_0(s/\langle s \rangle)$ shows reasonable agreement especially when $s/\langle s \rangle \geq 2$. Therefore, it is possible that Al which has a high bonding energy to As is hard to migrate and thus worked as nucleation center for QD formation.

## IV. SUMMARY

Self-assembled InAlAs/AlGaAs quantum dots (QDs) were grown on GaAs by molecular beam epitaxy, with the Stranski-Krastanov mode. A volume scaling function was found for the first time in ternary alloy QDs. The scaling function was similar to that of InAs/GaAs QDs, which is close to the scaling function for a two-dimensional (2D) submonolayer homoepitaxy model with a critical island size of $i = 1$. However, a character of $i = 0$ was also found beyond $s/\langle s \rangle = 2$.

FIG. 1. (Color online) Structure of the epitaxial layer. The top QD layer was measured by AFM.

FIG. 2. (Color online) Diameter and height histograms of QDs grown at 480°C and 510°C. Inset numbers are relative coverage $t/t_{3D}$ of 1.0, 1.2, and 1.4. Solid curves are results of Gaussion fit.

FIG. 3. (Color online) Volume histograms of QDs grown at 480°C and 510°C. Inset numbers are relative coverage $t/t_{3D}$ of 1.0, 1.2, and 1.4.

FIG. 4. (Color online) (a) The average diameter and average height of QDs grown at 480°C and 510°C as a function of $t/t_{3D}$ from Ref. 18. (b) Fluctuation of height, diameter, and volume as a function of $t/t_{3D}$ at 480°C and 510°C.

FIG. 5. (Color online) Experimental scaling of QDs grown at 480°C and 510°C with $t/t_{3D}$ was varied from 1.0 to 1.5. The solid, dashed, and dotted curves are the scaling function $f_i(s/\langle s \rangle)$ of Eq. (2) with $i$ = 1, 2, and 3, respectively. There is no fitting parameter in these curves.

FIG. 6. (Color online) Average of scaling plot in Fig. 5. Ts of 480°C and 510°C are shown in solid circles and triangles, respectively. The dashed curve is extracted from the

$i = 0$ simulation result by J. G. Amar.[11] The solid curve is the scaling function $f_i(s/\langle s \rangle)$ of Eq. (2) with $i = 1$.

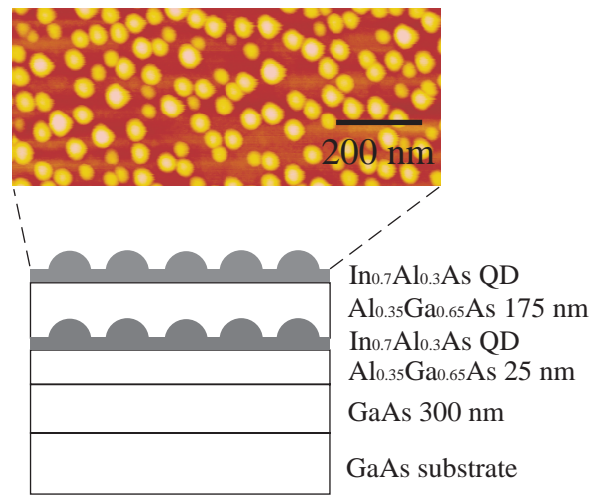

FIG. 1. (Color online) Structure of the epitaxial layer. The top QD layer was measured by AFM.

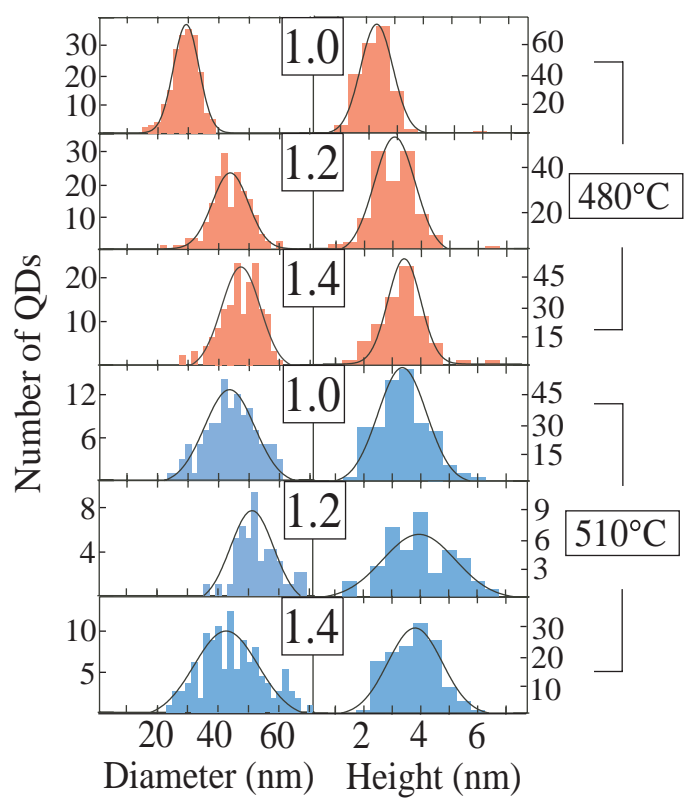

FIG. 2. (Color online) Diameter and height histograms of QDs grown at 480°C and 510°C. Inset numbers are relative coverage $t/t_{3D}$ of 1.0, 1.2, and 1.4. Solid curves are results of Gaussion fit.

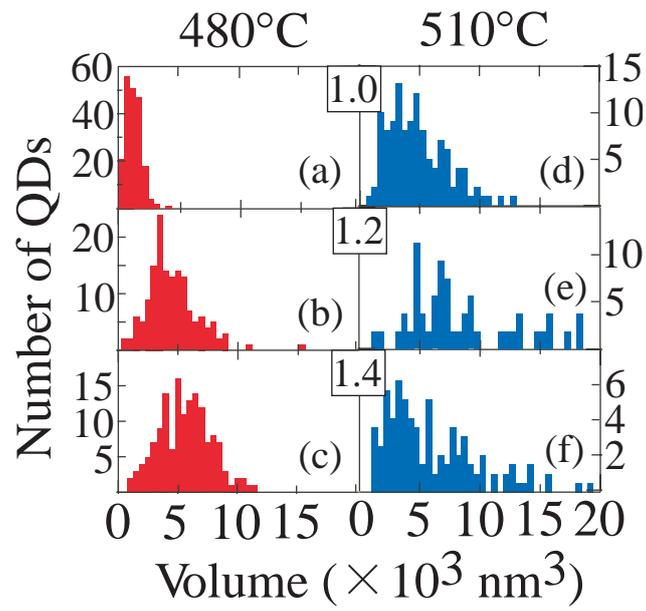

FIG. 3. (Color online) Volume histograms of QDs grown at 480°C and 510°C. Inset numbers are relative coverage $t/t_{3D}$ of 1.0, 1.2, and 1.4.

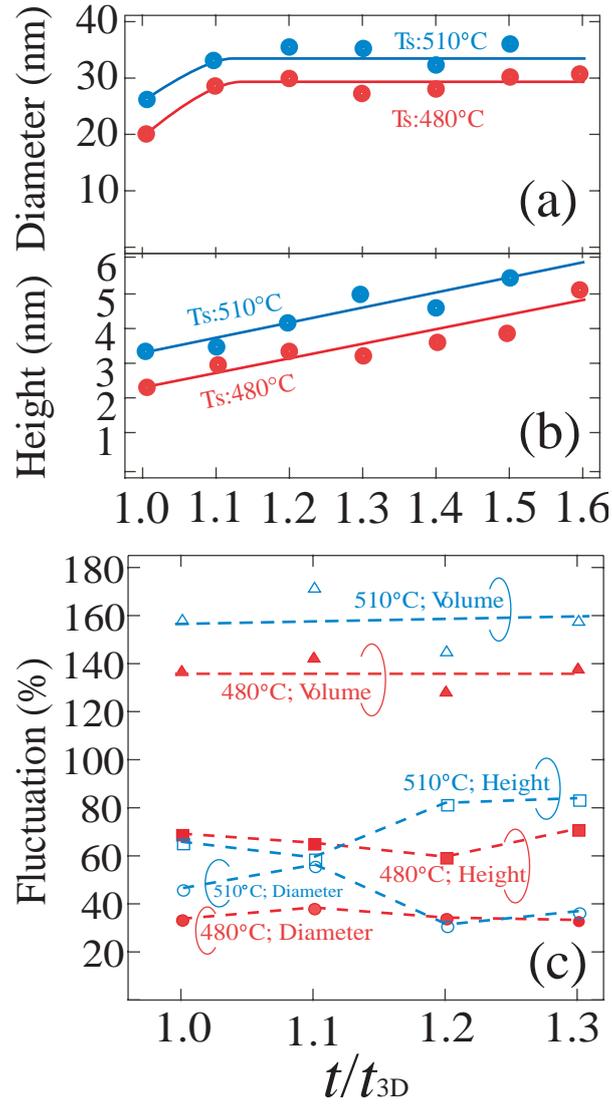

FIG. 4. (Color online) (a) The average diameter and average height of QDs grown at 480°C and 510°C as a function of $t/t_{3D}$ from Ref. 18. (b) Fluctuation of height, diameter, and volume as a function of $t/t_{3D}$ at 480°C and 510°C.

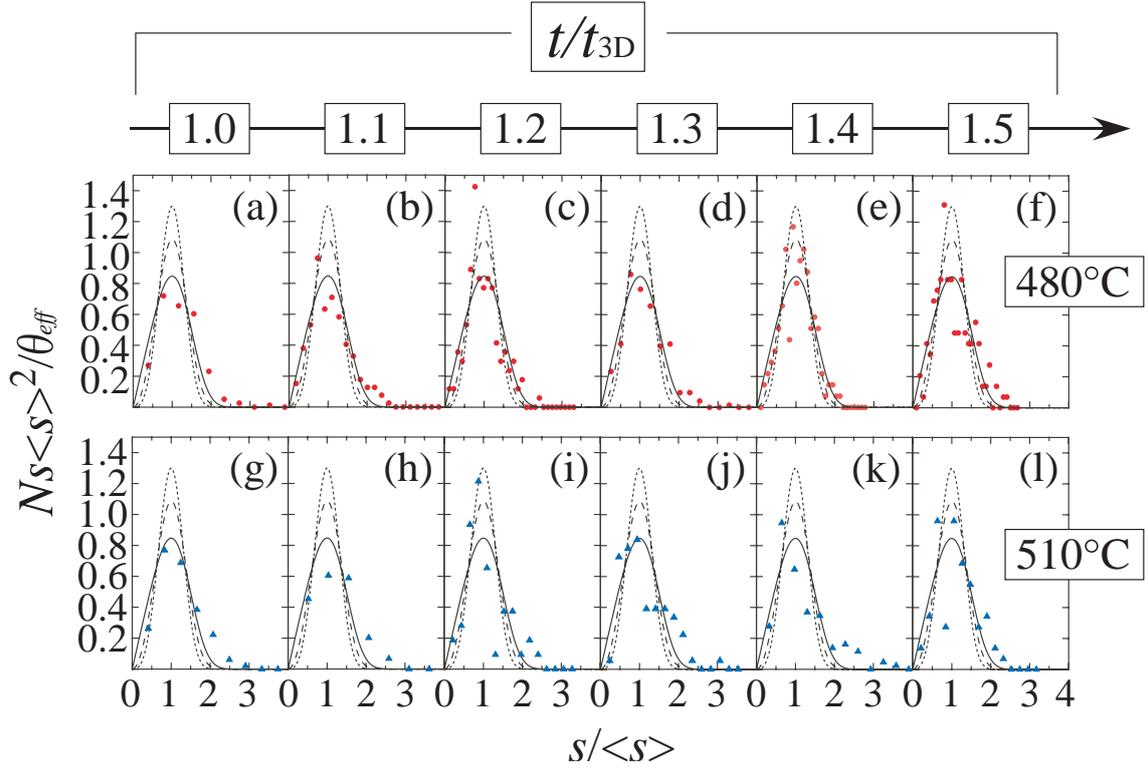

FIG. 5. (Color online) Experimental scaling of QDs grown at 480°C and 510°C with $t/t_{3D}$ was varied from 1.0 to 1.5. The solid, dashed, and dotted curves are the scaling function $f_i(s/\langle s \rangle)$ of Eq. (2) with $i$ = 1, 2, and 3, respectively. There is no fitting parameter in these curves.

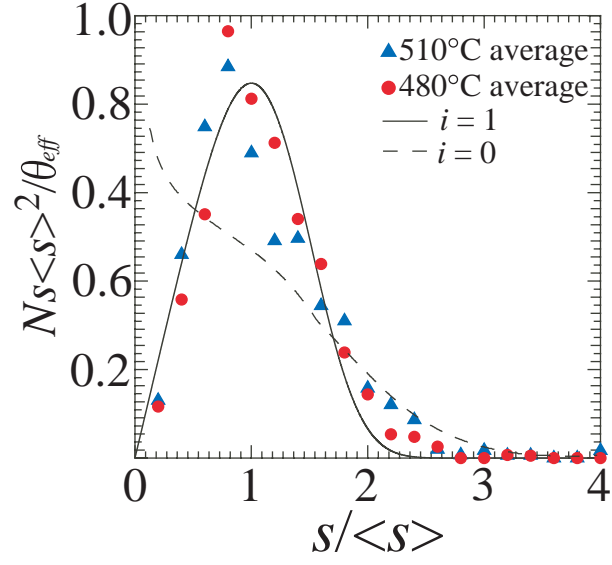

FIG. 6. (Color online) Average of scaling plot in Fig. 5. Ts of 480°C and 510°C are shown in solid circles and triangles, respectively. The dashed curve is extracted from the $i = 0$ simulation result by J. G. Amar.[11] The solid curve is the scaling function $f_i(s/\langle s \rangle)$ of Eq. (2) with $i = 1$.